\documentstyle[12pt]{article}
\makeatletter
\newcommand{\singlespacing}{\let\CS=\@currsize\renewcommand{\baselinestretch}
{1.0}\tiny\CS}
\newcommand{\doublespacing}{\let\CS=\@currsize\renewcommand{\baselinestretch}
{1.5}\tiny\CS}

\begin{document}
\title{The Angular-momentum aspect of {\it Shift} in FQHE }
\author{Dipti Banerjee \\
Department of Physics \\
Rishi Bankim Chandra College \\
Naihati,24-Parganas(N)\\
Pin-743 165, West Bengal\\
INDIA}
\date{}

\maketitle

\abstract
The angular momentum of the Hall particles acquiring
Berry phases from quantization,is the very source of
unifying the physics of IQHE and FQHE.
The origin of {\it Shift} quantum number lies in
the change of angular momentum of the Hall particles
from higher to lower
Landau level.
Here the non-vanishing effect of {\it Shift} in 2D has been
evaluated in the framework of $Z_p$ spin system and in view of these,
Ising model and $XY$ model are studied.

 \pagebreak

In a magnetic field ${\it B}$, when a particle is transported over
a closed contour, a non-integral phase is developed known as Dirac
phase that is determined by the flux through this contour[1].
\begin{equation}
\phi_D = \frac{e}{c\hbar}\int A^{\alpha}d
r_{\alpha}=\frac{e}{c\hbar}\int F^{\alpha\beta}dS_{\alpha\beta}=
\frac{e}{c\hbar}\int BdS = \frac{e}{c\hbar}\Phi
\end{equation}
This phase of Dirac is similar to the Berry phase [2] when the
eigenvectors of a particle are described by the parameter
dependent Hamiltonian. The nontrivial connection $A_\mu$ under the
parallel transport of the eigenvectors acquire the phase.
\begin{equation}
\phi_B= \oint A_\mu d\mu = \int F_{\mu\mu} ds
\end{equation}
 The curvature $F_{\mu\mu}$ associated with Berry connection
$A_\mu$ turns out to be nonzero in the adiabatic approximation and
occurs for loops in the $\mu$ space in the manner as the magnetic
field $B$ does. This gauge field is the origin of the Chern-Simons term
in both 2D and 3D cases. The effect of the Chern-Simons term is to
 associate with each particle a magnetic
 flux $e/\mu$. It may be viewed as if a certain number
  of flux quanta is attached to
 each electron leading to a gauge theoretic extension.This is the
 picture of composite particles in condensed matter Physics.
  This CS term itself generates a phase factor, in
 addition to the one generated by the Aharonov-Bhom factors
 associated with $\int d^3x{\vec{j}.\vec{A}}$ [3].
 There is a significant difference between the statistics of
 ordinary charge/flux tube composites, and that of objects where
 the charge-flux relation is determined dynamically by the equation
 of motion of the gauge field. In fact it is the induced charge
 that is proportional to the induced flux.
  $$\int q_{ind}(\Phi) d\Phi$$
 As a result the amount of charge
 present is less than the final amount and the resulting angular
 momentum is less than before.

This above analysis is reflected in the works of Bandyopadhyay[4] where
it has been shown that the charged particle in the $3D$ anisotropic
space (in an intense magnetic field) has the conserved angular momentum
\begin{equation}
\vec{J} = \vec{r}\times \vec{p}-\mu\vec{r}
\end{equation}
where $\mu$ is associated with the induced effect of flux
in the angular momentum.

In a Landau problem, when $2D$ electron gas
is taken to reside on the surface of a $3D$ sphere of large radius
R in a radial (monopole) strong magnetic field, particles
achieve quantization
acquiring the Berry phase through chiral symmetry breaking[6].
The situation arises when the system $\vec{H_o}$ interacts with
certain internal Hamiltonian $\vec{h_{int}}$ to have
$\vec{H}=\vec{H_o}+\vec{h_{int}}$. The response of the internal
system corresponding to the change of the external canonical
system becomes a part of the action as to associate the change of
fields by[7]
\begin{equation}
S_{eff}=S_o+ \hbar\Gamma(C)
\end{equation}
 The geometrical phase $\Gamma(C)$, is analogous to
the magnetic flux associated with the induced vector
potential $A=i<0|d|0>$ which emerges from Dirac pole.
This eventually causes the deformation of the symplectic structure and
the canonical variables originally defined are no longer
canonical.

 It has been pointed out by Bandyopadhyay [8] that the
 Berry phase may be associated with
 the deformation of the symplectic structure which can be
  linked up with the various
 quantization procedure. Indeed the relativistic generalization of
 the quantization procedure can be formulated in terms of gauge
 theoretic extension of the relativistic quantum particle.
 The inclusion of constraint of area preserving
diffiomorphisim in the Landau problem of $2D$ electron gas
 may be considered as the magnetic translations leading to
 the quantization of the system.
 This suggests that there is a dual relationship between the deformation
parameter $q$ of the symmetry group (algebra) with the Berry phase
parameter $\mu$ where the phase is given by $\exp{2\pi i \mu}$.
 Bandyopadhyay pointed out that the deformation through
 the anisotropy in space-time can be realized in any dimension,
 with the connections between the Chern-Simons term in the $(2+1)$ dimension,
 the chiral anomaly in $(3+1)$ and the conformal field theory in $1+1$-dimension.

In the light of Haldane [9], we first consider the statistical interaction
in $3+1$ dimension. Thus,the $2D$ Hall surface is a boundary
of a $3D$ sphere having radius R,in a radial (monopole) magnetic
field $B=\frac{\hbar S}{eR^2}$.
This $2S=N_\phi$ is an integer which defines the total
number of magnetic flux
through the surface. For parent state and the hierarchical state,
 the respective fluxes are
 $$S=1/2m(N-1)$$
 $$S=1/2m(N-1)\pm 1/2(\frac{N}{p}+1)$$
These show that the filling factor
 satisfies a slightly deviated relation as
\begin{equation}
2S=N_\phi={\nu}^{-1}N - {\cal S}
\end{equation}
Physicists identified this ${\cal S}$ as {\it Shift},
a topological quantum number developed due to
the coupling between the orbital spin and curvature of the space.
This orbital spin is
different from the spin of electrons and is associated with orbital
 angular momentum in cyclotron motion.
 In thermodynamic limit, this {\it Shift} becomes insignificant
leading the filling factor in each Landau
 level by $\nu=\frac{N}{2S}=\frac{N}{N_\phi}$.

In this anisotropic space, the chiral symmetry breaking
originates from the curvature $F_{\mu\nu}$ arises from induced flux
\begin{equation}
F_{\mu\nu}= \partial_\mu A_\nu -\partial_\nu A_\mu + [A_\mu,A_\nu]
\end{equation}
The vector potential $A_\mu\epsilon SL(2C)$ is responsible for
the deformation which can be linked up with the chiral anomaly.
This non-dynamical gauge $A_\mu$ is associated
with the flux which in $2+1$ dimension is the very cause of
 Chern-Simon term appeared in the Lagrangian.
\begin{equation}
L_{CS} = \frac{\mu}{2}\epsilon^{\mu\nu\lambda} A_\mu \partial_\nu A_\lambda
\end{equation}
This licenses a conservation of topological current $J_\mu$ which
include a topological invariant term in the $(2+1)$ dimension
\begin{equation}
 H = \frac{\theta}{2 \pi} \int d^3 x A_\mu J^\mu
\end{equation} in the action.
In fact it is the Hopf invariant describing basic maps of $S^3$ to
 $S^2$.If $\rho$ denotes a four dimensional index then we find [5]
 \begin{equation}
 \partial_\rho \epsilon^{\rho\mu\nu\lambda}A_\mu F_{\mu\nu}
 = \frac{1}{2}\epsilon ^{\rho\mu\nu\lambda}F_{\rho\mu}F_{\nu\lambda}
 \end{equation}
 which connects the Hopf invariant with chiral anomaly.This Hopf
 term plays a role somewhat similar to the role played by the
 Wess-Zumino interaction in
 connection with $3+1$ dim. Skyrmion term.

We have pointed out earlier[10] that the quasi-particles in the
hierarchies are formed when additional
flux is attached with quantized particles, which in the effective
theory is visualized through the additional CS terms in the Lagrangian on
(3+1)dimension.Thus quantization is the very cause of anisotropy on the Hall
surface along with the internal anisotropy of the particles.
 Apart from the gauge extension $A_\mu$ induced by the external strong
magnetic field strength $F_{\mu\nu}$, each particle have its own
internal gauge $C_\mu \epsilon SL(2C)$ which is visualized through
the field strength $\tilde{F_{\mu\nu}}$.
\begin{equation}
{\tilde F}_{\mu\nu} = \partial_\nu C_\mu - \partial
_\mu C_\nu + [ C_\mu,
C_\nu]
 \end{equation}
These two gauges $A_\mu$ and $C_\mu$
act as fibres at each point of base space occupied by Hall particles.
 Hence the quantization of particles in Hall hierarchies involve
  gauges which can be visualized through their change of
  coordinates from $x_\mu$ to $x_\mu+i(A_\mu+C_\mu)$.

 On the surface of 3D sphere the chiral symmetry breaking of
composite fermion in Hall fluid is associated with internal and external gauge
fields $F_{\mu\nu}$ and $\tilde{F_{\mu\nu}}$ and their physics can be
 described by the topological Lagrangian [10]
\begin{equation}
{\cal L}= -\frac{\theta}{16\pi^2} Tr^* F_{\mu\nu}F_{\mu\nu}
-\frac{\theta^\prime}{16\pi^2} Tr^* F_{\mu\nu}\tilde{F}_{\mu\nu}-
\frac{\theta^{\prime\prime}}{16\pi^2}
Tr^*\tilde{F}_{\mu\nu}\tilde{F}_{\mu\nu}
\end{equation}
Here every term corresponds to a total divergence of a topological
quantity, known as Chern-Simons secondary chracteristics class
defined by
\begin{equation}
{\Omega^\mu}_e =-\frac{1}{16\pi^2} \epsilon_{\mu\nu\alpha\beta}
Tr[A_\nu F_{\alpha\beta}-2/3(A_\nu A_\alpha A_\beta)]
\end{equation}

\begin{equation}
{\tilde{\Omega}^\mu} =-\frac{1}{16\pi^2}
\epsilon_{\mu\nu\alpha\beta} Tr[C_\nu F_{\alpha\beta}-2/3(C_\nu
A_\alpha A_\beta)]
\end{equation}

\begin{equation}
{\Omega^\mu}_i=-\frac{1}{16\pi^2}\epsilon_{\mu\nu\alpha\beta}
Tr[C_\nu \tilde{F}_{\alpha \beta} - 2/3(C_\nu C_\alpha C_\beta)]
\end{equation}

Assuming a particular choice of coupling $\theta=\theta'=\theta"$
in the Lagrangian the topological part of the action in (3+1)
dimension become
\begin{equation}
W_{\theta}=2(\mu_e + \mu_i + \tilde{\mu})\theta
\end{equation}
It gives rise the topological phase of Berry on the parallel
transport over a closed path of a Hall hierarchy state.
\begin{equation}
\phi_B = \pi W_\theta = 2\pi{\tilde{\mu}}_{eff}\theta = 2\pi(\mu_e
 + \mu_i + \tilde{\mu})\theta
 \end{equation}
 Here the first term $\mu_e$ is associated with Berry Phase (BP) factor
 due to the anisotropy in the external space.
  The second term gives
  rise to the inherent BP factor $\mu_i$ associated with
  the chiral anomaly of a free electron for its internal anisotropy
  and the third one $\tilde{\mu}$ effectively
  relates the coupling of the external
  field with the internal one.
   It forces the Hall particles in
   the presence of perpendicular magnetic field to
    move around the quantized circular orbit.
  The angular momentum of the rotating Hall particles varies
  with the radius of the orbit.Hence BP varies directly with
 angular momentum. With these views, we in our previous works
 are able to unify the physics of IQHE, FQHE
[5] through this angular momentum factor $\mu$ of the Hall
particles acquiring Berry Phase by
$2\pi\mu i$ having $\mu=0,\pm1/2,\pm1, \pm 3/2.....$
Further their hierarchies are studied [10]
 through the angular momentum aspect of Berry`s topological phase.
Hence the topological phase of the quantized Hall quasi-particles is
 \begin{equation}
 \phi_B = 2\pi({\mu}_{eff} +{\cal S})\theta
 \end{equation}
representing the solid angle subtended by their
 gauges (acting as vortex) with the axis of quantization about which
 the composite-particles precesses. This is a 3D manifestation of
 the topological phase.On 2D the Berry Phase is manifested through angles.
The {\it Shift} extends these angles by introducing some deformations.

It has been pointed out earlier[10] that the $\mu_{eff}$ appeared in topological phase (eqn.17)
actually visualizes the filling factor through the relation
$\nu=\frac{n'}{2\mu_{eff}}$, where $n'$
denotes the $n'th$ Landau level. In fact this $\mu_{eff}$ satisfies the
Dirac quantization condition
 \begin{equation}
 e^{\prime}\mu_{eff}=\pm \frac{n'}{2}
 \end{equation}   where $n'=1,2,3..$ denotes the hierarchy levels.
 Each quasiparticle in the $n'^{th}$ Landau level having charge
 $e^{\prime}$ behaves as a composite fermion.
 It will behave as fermion only when
 \begin{equation}
\tilde{e}(\mu_{eff}\mp\frac{n'\pm 1}{2})=\pm\frac{1}{2}
 \end{equation}
 Here the $+(-)$ sign indicates the orientation of the vortex
 line. The $(n'\pm1)/2$ is the magnetic strength $\mu$ of
 the additional quanta which is associated with the composite
 fermion in the higher Landau level to change it fermion in ground level.
For $\mu=\pm1/2,\pm 3/2,...$ the quanta behaves like fermion and
$\mu=\pm1,\pm2,....$ it shows bosonic behavior.Hence the behavior of
composite particle will be fermionic provided it follows
\begin{equation}
\tilde{e}\tilde{\mu}=\pm \frac{1}{2}
\end{equation}
having charge $\tilde{e}$ and flux $\tilde{\mu}$.
These composite particles satisfying eqn.(20) not always an
elementary fermion possessing unit electronic charge $(e=1)$
 and magnetic charge $(\mu=\pm1/2)$.
{\it Shift} is visualizing this change of magnetic charge by
 \begin{equation}
 2\tilde{\mu}={\cal S} = 2\mu_{eff}\mp(n'\pm 1)
 \end{equation}
In other words,
 \begin{equation}
{\cal S} = J_{CF}\mp(n'\pm 1)
 \end{equation}
 where the angular momentum of the composite fermion
 $J_{CF}$=$2\mu_{eff}$ is associated with the attached
 induced flux $\mu_e$ in addition with internal flux $\mu_i$.
It may be noted that the angular momentum of a
 quantized Hall particle is $J= \mu_i$
 and following Haldane [9]
 the Landau filling factor $\nu=1/m$ of parent state in
 the Hall fluid is the relative angular
 momentum $m=J_{ij}=J_i+J_j$ of two Hall particles [5].

With these ideas the filling factor of the composite particle is
 $$\nu_{CF}=\frac{n'}{2\mu_{eff}}=\frac{n'}{J_{CF}}$$
and leads us to realize ${\cal S}$ by
the fermionic filling factor whose relationship
with the composite fermionic behavior of the Hall particles
becomes.
 \begin{equation}
 {\nu_F}^{-1}=n'{\nu_{CF}}^{-1}\mp (n' \pm 1)
 \end{equation}
In 2D Hall surface,
Wen [11] pointed out vanishing effect of {\it Shift}. It seems to us
 that the effect of {\it Shift} will not vanish completely on
2D, since anisotropy incorporates some deformation of symplectic
structure on Hall surface visualized through curvature of space-time.
This physics can be recast through the representation of Hall particles
as $Z_p$ spin system.
Indeed, the $Z_p$ spin model can be well described by a system of scalar
particles, each having an orbital angular momentum $J=1/p$,
can form only one of the $p$ discrete angles $\theta_m=2\pi m/p$
 with some fixed direction in the space of
 internal degrees of freedom. This gives rise
 to the behavior of such particles having fractional statistics [12].
An equivalent relationship between
the anyon statistical formulation of the hierarchical states
depicted in the continued fraction scheme and Jain classification
scheme based on the composite fermion model has been derived [13]in the
background of $Z_p$ spin system.

The statistics of the particles $\theta$ in this $Z_p$ spin system is
related to the angular momentum by $J$[3].
\begin{equation}
\exp^{2\pi i J} = \exp^{i\theta}
\end{equation}
Here the statistical parameter is represented by
\begin{equation}
\theta= \pi[1-\frac{1}{n}]
\end{equation}
where $n=\infty$ corresponds to a fermion and $n=1$ a boson leading
to their respective statistics $\theta=\pi$ and $\theta=0$.
 We find using $J=1/p$ that $n$ is related to $p$ through the relation
\begin{equation}
n=\frac{p}{p-2}
\end{equation}
For $p=2$ and $n=\infty$ the system is
$Z_2$ equivalent to a fermionic gas. Also
for $p=3$ and $4$ results $n=3$ and $2$, respectively, the
system can be represented by fermions of fractional fermion
number.On the other hand as $p\longrightarrow \infty$ we have
$n=1$ system represents plane rotor model (XY model). This
analysis suggests that we can go gradually from the Ising model
($Z_2$) to the XY model through changing $p$ values.

As a fermion moving with
angular momentum $J=1/2$ departs from its pure fermionic character
to $Z_p$ spin system, it acquires the angular momentum $J=1/p$.This
leads to the change of angular momentum by
\begin{equation}
\Delta J= 1/2- 1/p= \frac{p-2}{2p}=\frac{1}{2n}
\end{equation}
 which further indicate a shift in statistics
\begin{equation}
\Delta \theta= 2\pi\Delta J
\end{equation}
If the composite fermion is in $Z_2$ spin
system, changes to fermion in the ground state, there exists
a shift in angular momentum by
$$ \Delta J=1/2-1/2 = 0$$ leading to change of statistics
$\Delta\theta=0$.
Similarly for the composite fermion in the XY model, the deviation of
angular momentum becomes
$$\Delta J=1/2-1/\infty = 1/2$$ resulting the corresponding
change in statistics of the Hall particles in the
XY model to fermionic model is $\Delta\theta=\pi$.

These deviation in angular momentum can be visualized
through the Shift quantum number in the frame work of
$Z_p$ spin system. Earlier we have studied [13] the
Hall particles in hierarchies by attachment of extra strength of vortex and
arrived at the Jain classification scheme from the
equivalent relationship between fractional statistics and
$Z_p$ spin system where the composite fermion filling factor is given by
\begin{equation}
{\nu_{CF}}^{-1}= \theta/\pi + \tilde{m}=\frac{2}{p}+ \tilde{m}=\frac{2mn
+1}{n}\cong\frac{2mn'\pm1}{n'}= 2\mu_{eff}/n'
\end{equation}
where $\tilde{m}$ is an even number in connection with the flux attached
with the composite particles.

In the light of $Z_P$ spin system, the {\it Shift}
in hierarchies from eqn.21 becomes
\begin{equation}
{\cal S}= n'(\frac{2}{p}+\tilde{m})\mp(n'\pm 1)
\end{equation}
in which using eqn. 27, for the release of the quanta the {\it Shift} is
\begin{equation}
{\cal S}= n'(\frac{2}{p}+\tilde{m})-(n'\pm 1)=
2n'\Delta J + 2mn'\pm 1\\
= 2n'\Delta J + n'{\nu_{CF}}^{-1}= 2n'\Delta J + J_{CF}
\end{equation}
On the other hand during acceptance of quanta, it becomes
\begin{equation}
{\cal S}= n'(\frac{2}{p}+\tilde{m})+(n'\pm 1)=
2n'(\Delta J + 1)+ 2mn'\pm 1 = 2n'(\Delta J + 1) + J_{CF}
\end{equation}
It seems from above, that {\it Shift} in both the cases depend
 on angular momentum of composite fermion and it's change.
The {\it Shift} during acceptance of quanta is more
than that of release, by the term $2n'$ which yields no extra effect
in the appearance of phase as seen from eqn.17.

A non-vanishing effect of {\it Shift}
in $2D$ is obtained
for its origin lying in the deviation of angular momentum
of the Hall particles from the higher to lower Landau level.
We further realize that the difference in angular momentum or the change in
statistics is the very source of the effective flux
$N_\phi$ (eqn5) present in any Landau level
by either
\begin{equation}
 2n'\Delta J~~~or~~~2n'(\Delta J + 1)
 \end{equation}
depending on the respective accept or release
of quanta by the Hall particles.

If the Hall fluid represent a plane rotor model(XY model), for
$p\longrightarrow \infty$ in the $Z_p$ system the ${\cal S}$ becomes
from eqn.(30)
\begin{equation}
{\cal S}= (2mn'\pm 1)\mp n'= n'{\nu_{CF}}^{-1}\mp n'= J_{CF}\mp n'
\end{equation}
which indicates $\triangle J=1/2$ and $\triangle\theta=\pi$.

On the other hand for Hall system behaving an Ising model,the {\it Shift}
results
\begin{equation}
{\cal S}=n'{\nu_{CF}}^{-1} + n'\mp n'
\end{equation}
following proper physics for
the change in angular momentum $\triangle J=0$ ,$\triangle\theta=\pi$
For acceptance or release of quanta the respective values of ${\cal S}$ are
\begin{equation}
{\cal S}=2mn'\pm 1 = n'{\nu_{CF}}^{-1} = J_{CF}~~~~ and
 {\cal S}= n'{\nu_{CF}}^{-1} + 2n'= J_{CF}+ 2n'
 \end{equation}
imply identical effect since $2n'$ produces no extra effect in
the topological phase. The effective flux associated with the respective
XY model and Ising model becomes
$$n',3n'~~~and~~~0 , 2n'$$ indicating a successive occurrence with the change of
$p$ values in Hall system.

We conclude by mentioning that
we have given here a new definition of {\it Shift} from the point of view of
 angular momentum. It indicates the connection between
fermionic and composite fermionic behavior of Hall particles
in FQHE. A non-vanishing effect of {\it Shift} in 2D
is shown here through the consideration of Hall particles
as $Z_p$ spin system.
We have also shed light on the effective flux
associated with each Landau level.
 In addition we are successful in
studying the physics of Hall system behaving as
the Ising model and XY model through {\it Shift} quantum number
 and effective
 flux which re-open a new idea of FQHE from the topological
 aspect.

{\bf Acknowledgement}\\
 I like to express my gratitude specially to Prof. P. Bandyopadhyay and all the
authors cited in my references.

\vspace{3cm}

{\bf References}
\begin{enumerate}
\item[1] K.Yu.Bliokh and Yu.P.Bilokh,quant-ph/0404144.
\item[2] M.V.Berry; Proc.Roy.Soc.London;{\bf A392}, 45 (1984),
Proceedings of a NATO Advanced Research
Workshop on Fundamental Aspects of Quantum Theory, held in Italy, Sep. 1985;
\item[3] F.Wilczek;{\it Fractional Statistics,Anyon Superconductivity},
 World Scientific,1990.
\item[4] P.Bandyopadhyay;{\it Geometry,Topology and Quantization}(kluwer,
 Dordrecht, The Netherlands,1996).
\item[5] D.Banerjee and P.Bandyopadhyay;Mod.Phys.Lett-{\bf B8},(1994),1643.
\item[6] D.Banerjee; Fort.Der.Physik{\bf 44}(1996)323-370.
\item[7] H.Kuratsuji and S.Iida; Phys.Rev {\bf D37},(1988),441.
\item[8] P.Bandyopadhyay;Int.J.Mod.Phys-{\bf A15},(2000),1415.
\item[9] F.D.M.Haldane, Phys.Rev.Lett {\bf 51}, 605 (1983).
\item[10] D.Banerjee; Phys.Rev.{\bf B58},(1998),4666.
\item[11] X.G.Wen; Adv.in Phys. {\bf 41},(1995) 405.
\item[12] G. Goswami and P.Bandyopadhyay;J.Math.Phys.{\bf 33}(1992),1090.
\item[13] D.Banerjee and P.Bandyopadhyay; Phys. Lett-{\bf A246},
(1998)181.
\end{enumerate}
\end{document}